\documentclass[%
reprint,
superscriptaddress,
amsmath,amssymb,
aps,
prl,
floatfix,
]{revtex4-1}

\usepackage{graphicx}% Include figure files
\usepackage{dcolumn}% Align table columns on decimal point
\usepackage{bm}% bold math
\usepackage{amsmath}
\usepackage{mathrsfs}
\usepackage{amssymb}
\usepackage{bm}
\usepackage{setspace}
\usepackage{array}
\usepackage{multirow}
\usepackage{float}
\usepackage{flushend}
\usepackage[pdfstartview=FitH,
CJKbookmarks=true,
colorlinks,
linkcolor=blue,
anchorcolor=blue,
citecolor=blue,
urlcolor=blue,
]{hyperref}
\usepackage{braket}

\begin{document}

\preprint{APS/123-QED}

\title{Simulation of Dynamical Quantum Phase Transition of the 1D Transverse Ising Model with a Double-chain Bose-Hubbard model}
\author{Ren Liao}
% \email{liaoren@pku.edu.cn}
\affiliation{School of Electronics Engineering and Computer Science, Peking University, Beijing 100871, China}
\author{Fangyu Xiong}%
% \email{xiongfangyu@pku.edu.cn}
\affiliation{Yuanpei College, Peking university, Beijing 100871, China}
\author{Xuzong Chen}%
\email{xuzongchen@pku.edu.cn}
\affiliation{School of Electronics Engineering and Computer Science, Peking University, Beijing 100871, China}

\begin{abstract} 
 We propose a spinless Bose-Hubbard model in an one-dimensional (1D) double-chain tilted lattice at unit filling per cell. A subspace of this model can be faithfully mapped to the 1D transverse Ising model through superexchange interaction with second-order perturbation theory. At a valid parameter region, numerical results show  good agreement of these two models both on energy spectrums and correlation functions. And we show that the dynamical quantum phase transition of the effective 1D transverse Ising model can be simulated. With carefully designed procedures for producing the dynamical quantum phase transition of the 1D transverse Ising model from a Mott insulator, the rate function of the recurrence probability to the ground-state manifold shows the same nonanalyticality at periodic time points as theory predicts. Our results may give some inspirations on simulating 1D transverse Ising model with superexchange interaction and exploring its dynamical quantum phase transition in experiment.  

  \begin{description}
    %   Secondary publications and information retrieval purposes.
    \item[PACS numbers]
      67.85.-d, 75.10.Hk
  \end{description}
\end{abstract}
\maketitle

Recently, enormous progress in simulating various kinds of quantum magnetism in cold atom systems opens new fascinating prospects for studying many interesting properties of magnetic models. However, experimental simulation of such magnetic models strongly depends on how the desired magnetic models are constructed in experiment. In a system with long-range interaction, the major difficulty lies on how spins are represented and how to manipulate the magnetic interaction between these spins precisely. And it is reported that the  transverse Ising model has been successfully implemented with ion trap\cite{Friedenauer2008,Kim2010,Kim2011,Lanyon2011,Britton2012}, Rydberg atom\cite{Elmer2018,Labuhn2016,Schauss2018} and superconducting quantum circuits\cite{Salathe2015,Gong2016,Barends2016,Harris2018}. And XXZ model has been realized with ultracold dipolar gases\cite{Hazzard2013,Paz2013}.  While in neural atom systems, it is primarily hindered by the magnetic interaction. In such systems it usually requires a special design of experimental conditions to realize localized spin representation and a strong enough magnetic interaction simultaneously. Several designs have been carried out in a two-component Fermi-Hubbard model~\cite{Russell2015,Laurence2016,Martin2016,Mazurenko2017,Drewes2017}, a spinless Bose-Hubbard model in a single tilted chain \cite{Greiner2011} and in a triangular lattice with shaking lattice~\cite{Struck2011} according to the propositions \cite{Efstratios1991,Sachdev2002,Eckardt2010} respectively. 

Despite these remarkable achievements, simulation of Ising model through superexchange interaction has not been reported. The difficulty lies on the contradiction of realizing an Ising-like spin-spin inteaction and the requirement of localized spin representation\cite{Lukin2003}. In this letter, we propose a model in a tilted double-chain lattice which can perfectly simulate the 1D transverse Ising model. In such a lattice, the nearest-neighbor superexchange interaction is modified by the tilting lattice while keeping atoms localized, making it possible to simulate an Ising-like spin-spin interaction.  Another concern is that superexchange interaction is usually vey weak comparing with typical accessible temperature in cold atom experiments. For example, a temperature of $k_BT\lesssim 4t^2/U$ is required for experimental observation of the magnetic order of the Fermi-Hubbard model at half-filling \cite{Mazurenko2017}. Here we show that dynamical quantum phase transition (DQPT) or quench dynamics could be used as another tool for revealing such weak magnetic orders. For the 1D transverse Ising model, DQPT requires a minimal spin-spin interaction in the order of $J_z\cdot t\sim 1$ (we set $\hbar=1$), where $t$ is the evolving time after quenching $h_x$ \cite{Heyl2013}. It is feasible to satisfy this requirement in experiment. Therefore, it is possible to explore weak magnetic orders induced by superexchange interaction through DQPT in neural atom systems. 
%For a typical superxchange interaction of tens of or hundreds of Hertz,
\begin{figure}[htbp]
  \includegraphics[width=\linewidth]{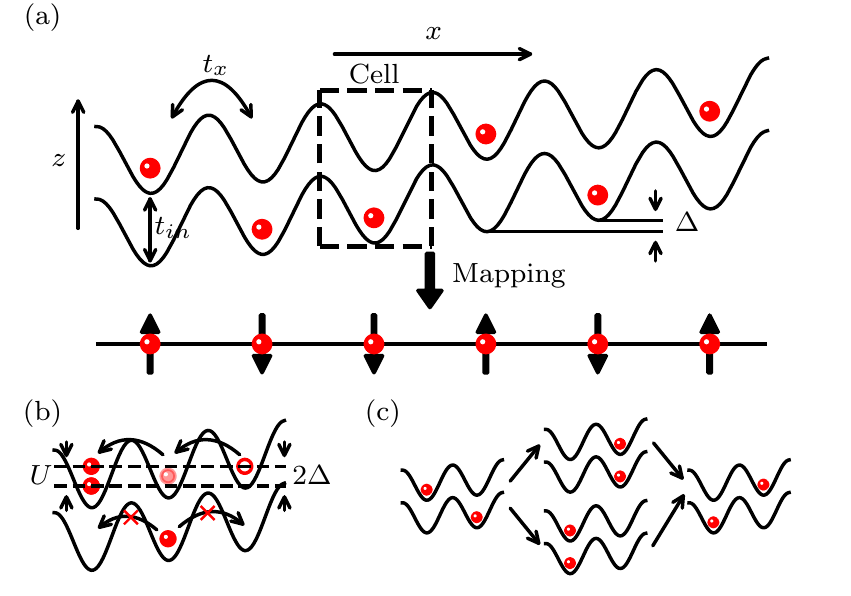}
  \caption{(a) Spin mapping of the half-filled Bose-Hubbard model in a double-chain tilted lattice. Spin up (down) is represented by the occupation of a spinless boson at the upper (lower) site in each cell. (b)  Localization of atoms in each cell is provided by the nearest lattice gap when $|\varDelta-U|\gg t_x, \varDelta\gg t_x$ and avoiding some other resonant points, such as the second-order resonant point $\varDelta=U/2$.  (c) When two atoms in nearest cells exchange their poistions through the above paired virtual hopping processes, the second-order superexchange interactions from these two processes cancel out totally. Only when these two atoms staying at the upper or lower site in each cell at the same time,  there is a nonzero second-order nearest-neighbor superexchange interaction. This means only an effective $S_i^zS_{i+1}^z$  spin-spin interaction exists in this model (see Supplementary Materials).}
  \label{fig:theory}
\end{figure}

\textit{Model and Methods.---} Our model begins with localized spin representation of the lattice model. As shown in FIG.~\ref{fig:theory}(a), spin up (down) is represented by the occupation of a spinless boson at upper (lower) site in each cell. Every atom is localized in a single cell (FIG.~\ref{fig:theory}(b)) because of the energy gap between nearest sites along x-axis when $|\varDelta-U|\gg t_x, \varDelta\gg t_x$ and avoiding some other resonant points. Here $U$ is the on-site interaction, $\varDelta$ and $t_x$ are the energy gap and the tunneling energy between two nearest sites along x-axis. All the states with only one atom per cell in the tilted 1D lattice form a subspace (denoted as $\mathscr{H}_\mathrm{one}$ hereafter) which can be mapped to the Hilbert space of a spin-1/2 Ising model $\mathscr{H}_\mathrm{Ising}$. 

Assuming a single-band case, this model can be described by a Bose-Hubbard model 
\begin{align}
\hat{H}=&\sum_{i,\sigma=\uparrow,\downarrow}\left[-t_x(\hat{c}_{i\sigma}^{\dagger}\hat{c}_{i+1\sigma}+h.c.)+\frac{U}{2}\hat{n}_{i\sigma}(\hat{n}_{i\sigma}-1)\right] \notag\\
&+\sum_{i}\left[i\varDelta(\hat{n}_{i\uparrow}+\hat{n}_{i\downarrow})+\frac{\delta_z}{2}(\hat{n}_{i\uparrow}-\hat{n}_{i\downarrow})\right]\notag\\
&-t_{in}\sum_i(\hat{c}_{i\uparrow}^{\dagger}\hat{c}_{i\downarrow}+h.c.)
\label{eqn:Hamiltonian}
\end{align}
where $t_{in}$ is a small tunneling energy inside each cell and $\delta_z$ is a small energy offset between the upper chain and the lower chain. Here $\uparrow$ and $\downarrow$ represent the upper and lower chain respectively. This model is isomorphic to a two-component Bose-Hubbard model in a single-chain tilted lattice with $U_{\uparrow\uparrow}=U_{\downarrow\downarrow}=U,  U_{\uparrow\downarrow}=0$. Applying the second-order perturbation theory, the effective Hamiltonian defined in the subspace $\mathscr{H}_\mathrm{one}$ can be written as
\begin{align}
  \hat{H}_{\mathrm{eff}}=\sum_iJ_z\hat{S}_i^z\hat{S}_{i+1}^z+h_x\hat{S}_i^x+h_z\hat{S}_i^z
  \label{eq:H_eff}
\end{align}
with
\begin{align}
J_z=\frac{8Ut_x^2}{\varDelta^2-U^2}, h_x=-2t_{in}, h_z=\delta_z.
\end{align}
 Here $\hat{S}_i^{\alpha}=\frac{1}{2}(\hat{c}_{i\uparrow}^{\dagger}\;\hat{c}_{i\downarrow}^{\dagger})\sigma^{\alpha}
(\hat{c}_{i\uparrow}\;\hat{c}_{i\downarrow})^\mathrm{T}$, $\alpha=x, y, z, \sigma^{\alpha}$ are Pauli matrices. Above deduction  is under the approximation  $\langle\hat{n}_{i\uparrow}+\hat{n}_{i\downarrow}\rangle=1$ for each cell corresponding to $\mathscr{H}_\mathrm{one}$.

\begin{figure}[!hb]
  \includegraphics[width=\linewidth]{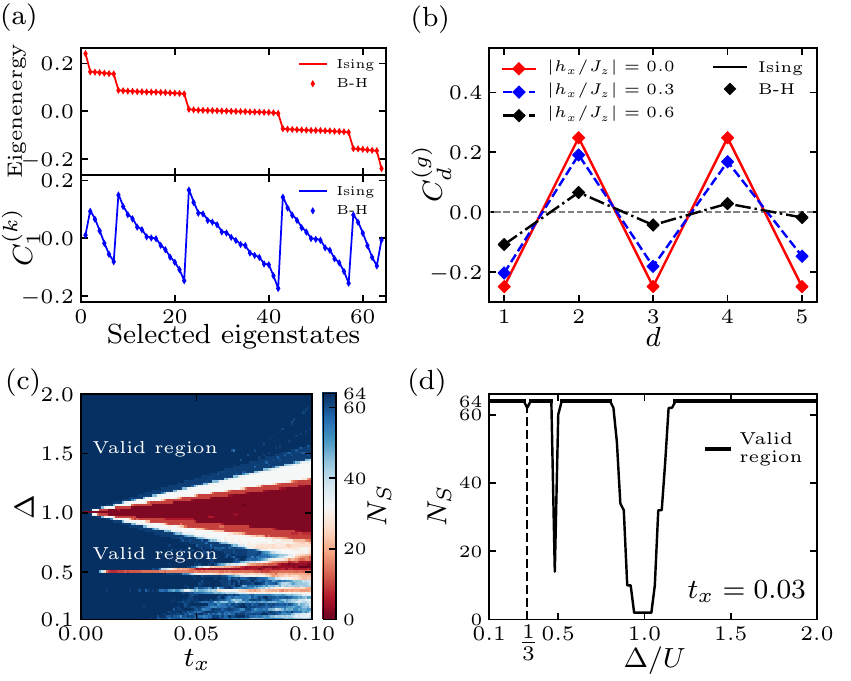}
  \caption{\textbf{Validation of mapping to the effective transverse Ising model.} (a) The energy spectrum and nearest-neighbor correlation functions $C_1^{(k)}$ of the selected eigenstates are both consistent with those of the effective transverse Ising model. The parameters are set as $ U = 1 $, $ \varDelta = 1.6 $, $ t_x = 0.04 $, $ t_{in} = 0.04 $, $ \delta_z = 0.01 $. (b) The correlation function of the ground state of the selected eigenstates $C_d^{(g)}$ at different $t_{in}$. The other parameters are the same as those in (a). It reveals antiferromagnetism of this state at specific parameters which also agrees with the effective transverse Ising model. (c)-(d) Number of selected eigenstates $N_S$ with respect to $\varDelta$ and $t_x$ at $ U = 1 $, $ t_{in} = 0 $ and $ \delta_z = 0 $. The valid region of parameters is in dark blue where $N_S=2^6$. Those peaks centered at $\varDelta/U=1/n (n=1,2,3,\cdots)$ signify $n^{\mathrm{th}}$-order resonant points where the mapping between these two models fails. }
  \label{fig:vadility}
\end{figure}

\textit{Validation of mapping to the 1D transverse Ising model.--- }To verify the validity of the effective transverse Ising model, an exact-diagonalization calculation is performed on the upper lattice model with six bosons in a $2 \times 6$ lattice on account of the limit of our computer. To find the subspace corresponding to $\mathscr{H}_\mathrm{one}$, we calculate the expectation value of $\langle\hat{n}_{i\uparrow}+\hat{n}_{i\downarrow}\rangle$ of each cell for each eigenstate and select those eigenstates satisfying  $\sum_{i=1}^L|\langle\hat{n}_{i\uparrow}+\hat{n}_{i\downarrow}\rangle-1|/L<\epsilon$ where $\epsilon$ is a small quantity (see Supplementary Materials). If all the parameters are set properly, the number of selected eigenstates $N_S$ is exactly $2^L$ which is just the size of the spin-1/2 model's Hilbert space. To make quantitative comparison between these two models, the energy spectrum and correlation functions of these selected eigenstates are calculated. For a lattice with finite size $L$, the correlation function $\langle\hat{S}_i^z\hat{S}_{i+d}^z\rangle$ for the $k$-th selected eigenstate $|\phi_k\rangle$ is defined as
\begin{equation}
  C_d^{(k)} =\sum_{i=1}^{L-d}\frac{\langle\phi_k|(\hat{n}_{i \uparrow}-\hat{n}_{i \downarrow})(\hat{n}_{i+d,\uparrow}-\hat{n}_{i+d,\downarrow})|\phi_k\rangle}{4(L-d)}.
\end{equation}
The results are depicted in FIG.~\ref{fig:vadility}(a) which shows very good consistency of these two models. And from the correlation function of the ground state of the selected eigenstates $C_d^{(g)}($Fig.~\ref{fig:vadility}(b)), we could see the antiferromagnetism of this state which also agrees with the effective transverse Ising model.

As demonstrated above, the validity of mapping the double-chain Bose-Hubbard model to the effective transverse Ising model depends on the localization of atoms, so that the spatial location of atoms in each cell can be mapped to a spin. However, there are some resonant points where atoms can hop to nearby cells by $n^{\mathrm{th}}$-order resonant tunneling (see Supplementary Materials). This will result in a number of selected eigenstates $N_S<2^L$. As shown in FIG.~\ref{fig:vadility}(c), the number of selected eigenstates $N_S$ regarding $\varDelta$ and $t_x$ is calculated at $U=1, t_{in}=0, \delta_z=0$. Beyond those valid regions where $N_S=2^L$ (here $L=6$), the $n$-th resonant points at $\varDelta=U/n (n=1,2,3,\cdots)$ can be determined by those peaks where $N_S$ drops below $2^L$ (FIG.~\ref{fig:vadility}(d)). These resonant points split the parameter plane and give a restriction on the selection of parameters to assure $N_S=2^L$. And at the valid region of the parameter plane, the validity of mapping to the effective transverse Ising model is guaranteed naturally.             
         
\textit{Simulation of dynamical quantum phase transition of the effective 1D transverse Ising model.---} With the upper model, we demonstrate that the dynamical quantum phase transition(DQPT) of the 1D transverse Ising model can be simulated. The typical characteristic of DQPT is the emergence of periodic nonanalytic points of a rate function when the system quenches across a quantum phase transition point from an initial ground state~\cite{Heyl2013}. This phenomenon has been observed in an ion trap system\cite{Blatt2017} and in a superconducting qubit circuit\cite{Fan2019} by directly simulating a transverse Ising model. And in a topological system, DQPT appears as a sudden creation or annihilation of vortex pairs at critical time points\cite{Sengstock2018}. But DQPT by simulating an 1D transverse Ising model in ultracold neural atom systems has not been reported yet. 

\begin{figure}[ht]
  \includegraphics[width = \linewidth]{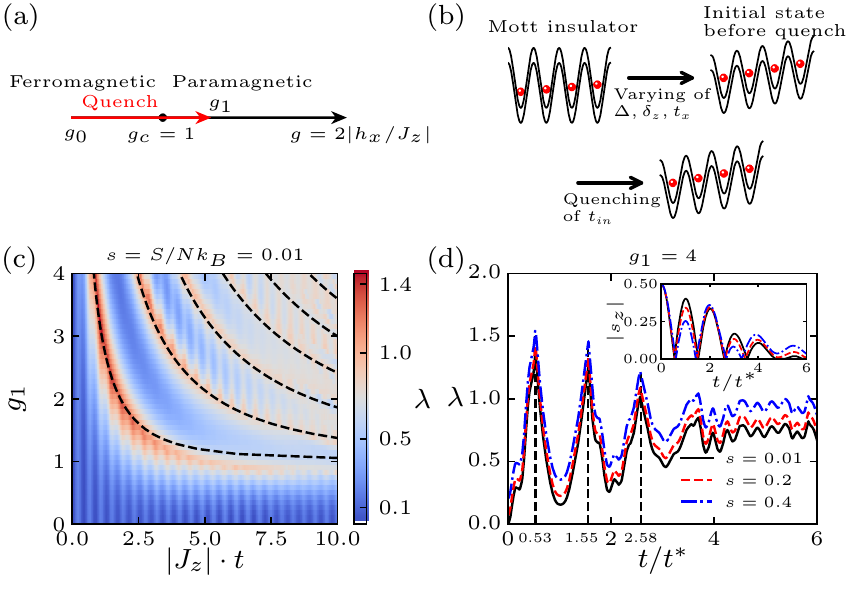}
  \caption{ \textbf{Simulation of the dynamical quantum phase transition of the effective 1D transverse Ising model.} (a) We use the quench scheme from $g_0=0$ to a designated $g_1$ to produce the DQPT of the effective transverse Ising model. (b) Procedures for producing the above quench scheme in the double-chain Bose-Hubbard model. The first step is to prepare an initial ferromagnetic state which is a ground state of the 1D ferromagnetic transverse Ising model at $g_0=0$. This can be realized by varying $\varDelta,\delta_z,t_x$ properly from an initial Mott insulator (see Supplementary Materials). Next the DQPT of the transverse Ising model can be produced by quenching $t_{in}$ from $t_{in}=0$ to a designated value. (c) The rate function $\lambda(t)$ simulated with the Bose-Hubbard model concerning different $g_1$ at an initial entropy $S/Nk_B=0.01$. The black dashed lines are theoretical results of the nonanalytic points of $\lambda(t)$ for $L\rightarrow\infty$ which shows very good consistency of these two models. (d) $\lambda(t)$ and $|s_z(t)|$ at $g_1=4$ regarding different initial entropy. $\lambda(t)$ becomes nonanalytical and $|s_z(t)|$ becomes zero at $t_n^*=(n+1/2)t^*$. And it can be noticed that the initial entropy has little influence on $\lambda(t)$ except for its magnitude, but has a noticeable effect on $|s_z(t)|$.}
  \label{fig:phase_transition}
\end{figure}

For dilute ultracold gas in optical lattice, it can be regarded as an isolated system if there is no obivious heating and atom loss within the time period of experiment. Thus, varying of system parameters preserves entropy and atom number. To simulate the DQPT of the effective transverse Ising model as shown of the red arrow in Fig.\ref{fig:phase_transition}(a), we first need to prepare an initial ground state of the Ising model at $J_z<0, h_z=0, h_x=0$, and then quench $t_{in}$ from $t_{in}=0$ to a designated value to produce the DQPT of the transverse Ising model. Here we use $g=2|h_x/J_z|=4t_{in}/|J_z|$ as the parameter to describe DQPT. The above initial ground state is a ferromagnetic state which can be produced from an initial Mott insulator by varying $\varDelta, \delta_z, t_x$ properly (Fig.\ref{fig:phase_transition}(b)). And the quench of $t_{in}$ can be simply implemented by a quench of lattice potential. However, DQPT requires a nearly pure quantum state as the inital state which means the initial state should be prepared with very low entropy. This is very difficult to realize in neural atom systems because of the weak interaction between atoms. But we show that this problem can be avoided in this model. Considering a real case with finite entropy which is assumed to be constant throughout the period of dynamical evolution, the system can be described by a density operator $\hat{\rho}(t)$ which follows the equation $i\frac{\partial\hat{\rho}}{\partial t}=[\hat{H}(t),\hat{\rho}]$. We assume the system is in thermal equilibrium in the beginning with $\hat{\rho}(0)=\operatorname{tr}(e^{-\hat{H}(0)/k_BT}/Z)$ to give the system an entropy, where the temperature $T$ is determined by the designated total entropy $S/Nk_B=-\mathrm{tr}(\hat{\rho}(0)\ln\hat{\rho}(0))/N$. Here $N$ is the total particle number.   $\hat{\rho}(t)$ can be decomposed as 
\begin{align}
&\hat{\rho}(t)=\hat{\rho}_{one}(t)+\hat{\rho}_{other}(t)\notag\\
&\hat{\rho}_{one}(t)=P_{+}(t)|+\rangle\langle+|+ P_{-}(t)|-\rangle\langle-|+\hat{\rho}_{one,else}
\end{align}
where $\hat{\rho}_{one}$ is the density operator in the subspace $\mathscr{H}_\mathrm{one}$. And $|+\rangle (|-\rangle)$ is the state with every atom on the upper (lower) site in each cell, corresponding to the two degenerate ground states of the transverse Ising model at $J_z<0, h_z=0, h_x=0$.  When $g$ is quenched from $g_0=0$ to a designated $g_1$, the rate function for such a small system can be introduced as \cite{Heyl2014,Bojan2018}
\begin{align}
  \lambda(t)=\frac{1}{L}\min(-\ln P_{+}(t),-\ln P_{-}(t)).
  \label{eqn:DQPT}
\end{align}

With proper approximation of the evolving process (see Supplementary Materials), the results are shown as FIG.~\ref{fig:phase_transition}(c) and FIG.~\ref{fig:phase_transition}(d). We can see the periodic nonanalytic behaviors of $\lambda(t)$ at certain times $t_n^*(g_0,g_1)=(n+1/2)t^*$. $t^*$ is approximately 1.03 times of the theoretical result $t^*(g_0,g_1)=\frac{2\pi}{|J_z|}\sqrt{(g_1+g_0)/(g_1-g_0)(g_1^2-1)}\;(g_1>1)$ for $L\rightarrow\infty$ \cite{Heyl2013}, as shown of the black dashed lines in FIG.~\ref{fig:phase_transition}(b). And only when $g_1$ is quenched across the quantum phase transition point $g_c=1$, there are periodic nonanalytical points of $\lambda(t)$. And as depicted in the inlet of FIG.~\ref{fig:phase_transition}(c), the magnetization $|s_z(t)|=|\operatorname{tr}(\hat{\rho}(t)\sum_i\hat{S}_i^z/L)|$ also becomes zero at $t_n^*$ when the entropy is small. These are all in accordance with the DQPT of 1D transverse Ising model. The deviation from the ideal case is mainly due to a small lattice size. The influence of the total entropy of the initial Mott insulator is shown in FIG.~\ref{fig:phase_transition}(c). It can be noticed that the total entropy has little influence on $\lambda(t)$ except for its magnitude, but has a noticeable effect on $|s_z(t)|$. This is becasue the subspace $\mathscr{H}_\mathrm{one}$ is decoupled with all the other subspaces in terms of second-order coupling at the valid parameter region so that $\hat{\rho}(t)$ can be decomposed as  $\hat{\rho}(t)=\hat{\rho}_{one}(t)\otimes\hat{\rho}_{other}(t)$. Thus, the evolution of $\hat{\rho}_{one}(t)$ after quenching is mainly governed by $\hat{H}_{\mathrm{eff}}$ defined in Eqn.(\ref{eq:H_eff}) by $i\frac{\partial\hat{\rho}_{one}(t)}{\partial t}=[\hat{H}_{eff},\hat{\rho}_{one}(t)]$, producing a similar $\lambda(t)$ as long as $\hat{\rho}_{one}(t=0)\approx P_{+}(t=0)|+\rangle\langle+|$ or $P_{-}(t=0)|-\rangle\langle-|$ regardless of the initial total entropy. Such a $\hat{\rho}_{one}(t=0)$ can be easily obtained by producing an initial Mott insulator with a large enough negative or positive $\delta_z$. In this way, the requirement of preparing a pure state as the initial state to produce DQPT can be avoided. While $|s_z(t)|=|s_z(t)|_{one}+|s_z(t)|_{other}$, $|s_z(t)|\approx|s_z(t)|_{one}$ is only estsblished when $\hat{\rho}(t=0)\approx\hat{\rho}_{one}(t=0)$.

\textit{Summary and outlook.---}In summary, we have proposed a double-chain Bose-Hubbard model in an 1D tilted lattice. The low-energy spectrum of a subspace $\mathscr{H}_{one}$ can faithfully simulate an 1D transverse Ising model at the valid parameter region. Meanwhile, we design a process of simulating the dynamical quantum phase transition of the 1D transverse Ising model from a Mott insulator which could be produced with usual experimental methods. The nonanalyticality of a rate function introduced for a small system shows good agreement with theoretical predictions. And we find that the nonanalyticality of the rate function can be avoided to be affected by the initial entropy of the Mott insulator, thus making it possible for experimental observation. Recently, simulation of a two-component Bose-Hubbard model in a single titled chain with ${}^7Li$ atom has been reported  \cite{Ivana2019}, revealing the superexchange interaction in a tilted lattice. This experiment  strongly supports the feasibility of realizing the above double-chain Bose-Hubbard model with bosonic ${}^7Li$ atom. Our results may give some inspirations of following experiments.

We thank for Prof. Dingping Li and Fei Gao for some useful discussions in the theoretical part. This work is supported by the National Natural Science Foundation of China (Grants Nos. 91736208, 11920101004, 11334001, 61727819, 61475007), and the National Key Research and Development Program of China (Grant No. 2016YFA0301501).

\bibliographystyle{apsrev4-1}
%\bibliography{ref}

\providecommand{\noopsort}[1]{}\providecommand{\singleletter}[1]{#1}%

\flushend

%% Supplementary Materials
\clearpage
\onecolumngrid
\begin{center}
	\LARGE
	\textbf{Supplementary Materials}
\end{center}
\vbox{}
\renewcommand{\thefigure}{S\arabic{figure}}
\setcounter{figure}{0}  
\twocolumngrid
\section{Effective Hamiltonian}
To derive the effective Hamiltonian $\hat{H}_{\mathrm{eff}}$ (Eqn. (2) in the main text) defined on the subspace $\mathscr{H}_{\mathrm{one}}$, we rewrite the Hamiltonian of the double-chain Bose-Hubbard model as $\hat{H}=\hat{H}_0+\delta\hat{H}+\hat{V}$ where 
\begin{flalign}
\hat{H}_0=&\sum_{i\sigma=\uparrow,\downarrow}\left[\frac{U}{2}\hat{n}_{i\sigma}(\hat{n}_{i\sigma}-1)+ i\varDelta\hat{n}_{i\sigma}\right]+ U_{\uparrow\downarrow}\sum_i\hat{n}_{i\uparrow}\hat{n}_{i\downarrow}\notag\\
\delta\hat{H}=&-t_{in}\sum_i(\hat{c}_{i\uparrow}^{\dagger}\hat{c}_{i\downarrow}+h.c.)\notag
+\frac{\delta_z}{2}\sum_i(\hat{n}_{i\uparrow}-\hat{n}_{i\downarrow})\notag\\
\hat{V}=&-t_x\sum_{i\sigma=\uparrow,\downarrow}(\hat{c}_{i\sigma}^{\dagger}\hat{c}_{i+1\sigma}+h.c.)\notag\nonumber
\end{flalign}
The introduced term $U_{\uparrow\downarrow}\sum_i\hat{n}_{i\uparrow}\hat{n}_{i\downarrow}$ is due to a computing error which will be explained later. The second-order perturbative Hamiltonian on the subspace $P_0$ is $\hat{H}_{\mathrm{eff}}^0=E_0+P_0[\delta\hat{H}+\hat{V}(E_0-\hat{H}_0)^{-1}\hat{V}]P_0$, where $E_0=L(L+1)\varDelta/2$ and $P_0$ is a projection operator on the subsapce which is composed of eigenstates of $\hat{H}_0$ with an eigenenergy $E_0$. $P_0=P_{one}\oplus P_{else}$, where $P_{one}$ is the projection operator on the subspace $\mathscr{H}_{\mathrm{one}}$. If there is no resonant coupling between $P_{one}$ and $P_{else}$, $P_{one}$ can be regarded as an independent subspace and $\hat{H}_{\mathrm{eff}}^0$ can be written as $\hat{H}_{\mathrm{eff}}^0=\hat{H}_{\mathrm{eff}}^{one}+\hat{H}_{\mathrm{eff}}^{else}$. Thus, in the subspace $\mathscr{H}_{\mathrm{one}}$, the leading second-order Hamiltonian is
\begin{equation}
\hat{H}_{\mathrm{eff}}^{one}=P_{one}[\delta\hat{H}+\hat{V}(E_0-\hat{H}_0)^{-1}\hat{V}]P_{one}.
\nonumber % '\nonumber' command cancels out the numbering of each formula
\end{equation} 
With $\hat{S}_i^{\alpha}=\frac{1}{2}(\hat{c}_{i\uparrow}^{\dagger}\;\hat{c}_{i\downarrow}^{\dagger})\sigma^{\alpha}(\hat{c}_{i\uparrow}\;\hat{c}_{i\downarrow})^\mathrm{T}$, $P_{one}\delta\hat{H}P_{one}$ is mapped to $\sum_i(-2t_{in}\hat{S}_i^x+\delta_z\hat{S}_i^z)$. And the second term can be expanded as  
\begin{align}
P_{one}\hat{V}(E_0-\hat{H}_0)^{-1}\hat{V}P_{one}=\sum_{\alpha,\beta,\gamma}|\alpha\rangle\langle\beta|\frac{\langle\alpha|\hat{V}|\gamma\rangle\langle\gamma|\hat{V}|\beta\rangle}{E_0-E_{0\gamma}}
\nonumber          
\end{align}
with $|\alpha\rangle,|\beta\rangle\in\mathscr{H}_{\mathrm{one}}$. These terms can be interpreted as two-step hopping processes, and they give rise to a superexchange interaction $\sum_i (\frac{8Ut_x^2}{\varDelta^2-U^2}-\frac{4U_{\uparrow\downarrow}t_x^2}{\varDelta^2-U_{\uparrow\downarrow}^2})\hat{S}_i^z\hat{S}_{i+1}^z+ \frac{4U_{\uparrow\downarrow}t_x^2}{\varDelta^2-U_{\uparrow\downarrow}^2} (\hat{S}_i^x\hat{S}_{i+1}^x+\hat{S}_i^y\hat{S}_{i+1}^y)$ given the restriction of $\hat{n}_{i\uparrow}+\hat{n}_{i\downarrow}=1$ for $\mathscr{H}_{\mathrm{one}}$. And when $U_{\uparrow\downarrow}=0$, $\hat{H}_{\mathrm{eff}}^{one}$ will change into a transverse Ising model
\begin{align}
\hat{H}_{\mathrm{eff}}^{one}=\sum_i(\frac{8Ut_x^2}{\varDelta^2-U^2}\hat{S}_i^z\hat{S}_{i+1}^z-2t_{in}\hat{S}_i^x+\delta_z\hat{S}_i^z).
\notag
\end{align}
The computing error at $U_{\uparrow\downarrow}=0$ can be avoided by setting a tiny nonzero $U_{\uparrow\downarrow}$. This will not affect the above transverse Ising model so much. Hereafter $U_{\uparrow\downarrow}=0.02$ is set as default in the following calculation if not specified.

%To achieve a ferromagnetic Ising model, the best option is to choose a $U$ and $\varDelta$ with $U<0, \varDelta>|U|$. This can give a larger $|J_z|$ comparing with choosing $U>0, \varDelta<U$ where some resonant points at $\varDelta=U/n (n=2,3,\cdots)$ restricts a maximal $t_x$. 

\section{Selected eigenstates}
If the above mapping is valid, the subspace $\mathscr{H}_{\mathrm{one}}$ should be an independent subspace of the double-chain Bose-Hubbard model. Thus, when solving for eigenstates of the double-chain Bose-Hubbard model, there should be exactly $2^L$ eigenstates whose energy spectrum and wavefunctions are nearly the same as those of the effective transverse Ising model. To find out these eigenstates, we can calculate $\langle\hat{n}_{i\uparrow}+\hat{n}_{i\downarrow}\rangle$ for each site of each eigenstate considering $\langle\hat{n}_{i\uparrow}+\hat{n}_{i\downarrow}\rangle=1$ for each state in $\mathscr{H}_{\mathrm{one}}$. So we directly solve for the eigenstates of the double-chain Bose-Hubbard model by exact-diagonalization and select those eigenstates satisfying $\sum_{i=1}^L|\langle\hat{n}_{i\uparrow}+\hat{n}_{i\downarrow}\rangle-1|/L<\epsilon$ where $\epsilon$ is a small quantity and is usually set to 0.05 in our calculation. The numerical result is shown as Fig. S1(a). It can be seen that the number of selected eigenstates $N_S$ is exactly $2^L$ and the energy spectrum of these eigenstates is also consistent with that of the effective transverse Ising model (Fig. S1(b)). And for each selected eigenstate, its wavefunction is also mainly composed of the states of $\mathscr{H}_{\mathrm{one}}$ and consistent with the wavefunction of the corresponding eigenstate of the effective transverse Ising model.

%%Meanwhile, for any selected eigenstates, the wavefunction $|\psi_{selected}\rangle$ has a form
%\begin{align}
%|\psi_{selected}\rangle=\sum_{|\phi_{\alpha}\rangle\in\mathscr{H}_{\mathrm{one}}} c_{\alpha}|\phi_{\alpha}\rangle+\sum_{|\phi_{\beta}\rangle\notin\mathscr{H}_{\mathrm{one}}}d_{\beta}|\phi_{\beta}\rangle\notag
%\end{align} 
%with $\sum_{\alpha}|c_{\alpha}|^2\gg\sum_{\beta}|d_{\beta}|^2$ so that the second term can be ignored. And the coefficients $c_{\alpha}$ are also consistent with those of the eigenstates of the above transverse Ising model. Considering $\langle\phi_{\alpha}|(\hat{n}_{i\uparrow}+\hat{n}_{i\downarrow})|\phi_{\alpha}\rangle=1$ for any $|\phi_{\alpha}\rangle\in\mathscr{H}_{\mathrm{one}}$, $\sum_{i=1}^L|\langle\psi_{selected}|(\hat{n}_{i\uparrow}+\hat{n}_{i\downarrow})|\psi_{selected}\rangle-1|/L$ will be a small number. This is why we set the criterion $\sum_{i=1}^L|\langle\hat{n}_{i\uparrow}+\hat{n}_{i\downarrow}\rangle-1|/L<\epsilon$ to find these eigenstates. 

\begin{figure}[tb]
	\includegraphics[width =\linewidth]{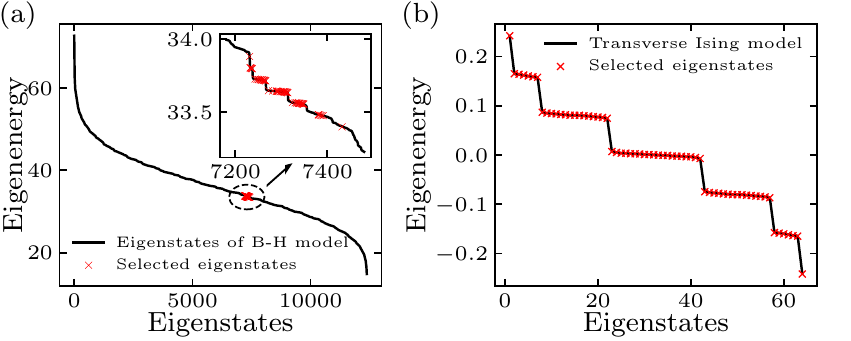}
	\caption{(a) Selected eigenstates shown in the eigenstates of the double-chain Bose-Hubbard model. The number of selected eigenstates satisfying $\sum_{i=1}^L|\langle\hat{n}_{i\uparrow}+\hat{n}_{i\downarrow}\rangle-1|/L<\epsilon$ is exactly $2^L$. (b) The energy spectrum of the selected eigenstates cut from (a) is highly coincident with that of the transverse Ising model, showing the validity of second-order perturbation theory. The parameters are set as $U=1, \varDelta=1.6, t_x=0.04, t_{in}=0.04,\delta_z=0.01,\epsilon=0.05,U_{\uparrow\downarrow}=0.02$.}
	\label{fig:figS1}
\end{figure}

\section{Resonant points}
The above deduction of $\hat{H}_{\mathrm{eff}}^{one}$ is invalid at some resonant points. At these points, $P_{one}$ and $P_{else}$ are coupled through $n^{\mathrm{th}}$-order ($n=1,2,3,\cdots$) resonant tunnelings. This will cause a failing of the independence of the subspace  $\mathscr{H}_{\mathrm{one}}$. Thus, when solving for eigenstates, the number of  selected eigenstates $N_S$ will be less than $2^L$. To reveal these resonant points, we calculate $N_S$ with respect to $U$ and $U_{\uparrow\downarrow}$ while keeping $\varDelta$ fixed. From Fig. S2(a), we could clearly infer these resonant points and their order from those lines where $N_S<2^L$. Generally a lower-order resonant point causes a wider area of $N_S<2^L$. And these resonant points have a general form $aU_{\uparrow\downarrow}+bU+c\varDelta=0$, where $a,b,c$ are small integers. At each rosonant point, there is a resonant coupling between a state $|\phi_{\alpha}\rangle\in\mathscr{H}_{\mathrm{one}}$ and a state $|\phi_{\beta}\rangle\notin\mathscr{H}_{\mathrm{one}}$, such as the third-order resonant point $2U_{\uparrow\downarrow}+U-\varDelta=0$ depicted in Fig. S2(b). By selecting $U, U_{\uparrow\downarrow}, \varDelta$ properly, we can avoid these resonant points to guarantee the validity of mapping to the above transverse Ising model.
\begin{figure}[tb]
	\includegraphics[width =\linewidth]{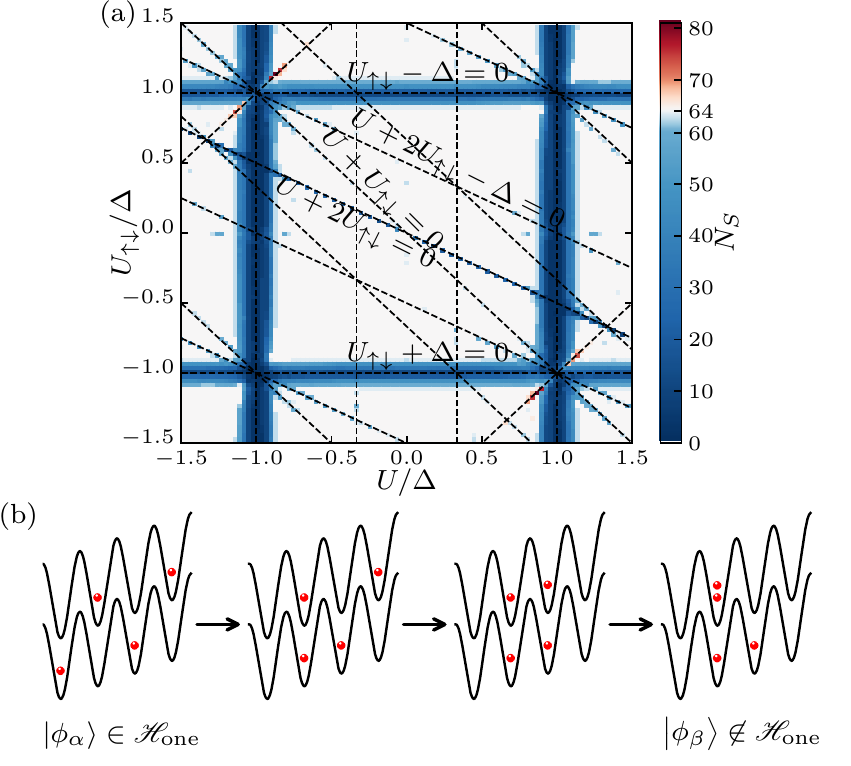}
	\caption{(a) The number of slected eigenstates $N_S$ with respect to $U$ and $U_{\uparrow\downarrow}$. Those lines with $N_S<2^L$ signifies the resonant points except the point at $U_{\uparrow\downarrow}=0$. The parameters are set as $\varDelta=1.5, t_x=0.04, t_{in}=0, \delta_z=0$. (b) At each resonant point in (a), there is a resonant coupling between a state $|\phi_{\alpha}\rangle\in\mathscr{H}_\mathrm{one}$ and a state $|\phi_{\beta}\rangle\notin\mathscr{H}_\mathrm{one}$, such as the third-order resonant point at $2U_{\uparrow\downarrow}+U-\varDelta=0$ in a four-cell lattice shown as above. This kind of resonant coupling will destroy the independence of the subspace $\mathscr{H}_\mathrm{one}$, which will make the mapping to the above effective Ising model invalid.}
	%Meanwhile, it can be verified that $U_{\uparrow\downarrow}=0$ is also not a third order or higher order resonant point.
	\label{fig:figS2}
\end{figure}

\section{The computing error at $U_{\uparrow\downarrow}=0$}
From Fig. S2(a), we could notice that there are some points with $N_S<2^L$ at $U_{\uparrow\downarrow}=0$. This error is presumably believed to be caused by the second-order resonant coupling shown in Fig. S3(a). However, the second-order superexchange interaction from the paired resonant tunneling processes from Fig. S3(a) cancels out totally, so that $U_{\uparrow\downarrow}=0$ is not a second-order resonant point, neither a high-order resonant point. We finally find this is caused by a computing error when solving for eigenstates. The computer program can not distinguish two degenerate states and there is a freedom in their coefficients, leading to a virtual coupling between these two degenerate states. Here the degeneracy is in terms of $\hat{H}_0$. This error sometimes occurs at $U_{\uparrow\downarrow}=0$. Therefore, we set a tiny nonzero $U_{\uparrow\downarrow}$ to lift the degeneracy between those states, such as the two states $|\phi_{\alpha}\rangle\in\mathscr{H}_\mathrm{one}$ and $|\phi_{\beta}\rangle\notin\mathscr{H}_\mathrm{one}$ in Fig. S3(a), to solve this problem. The result is shown in Fig. S3(b). The validity of $U_{\uparrow\downarrow}=0$ means that the transverse Ising model can be fully achieved in this way.

%generating some eigenstates containing terms like $u|\phi_{\alpha}\rangle+v|\phi_{\beta}\rangle, |\phi_{\alpha}\rangle\in\mathscr{H}_\mathrm{one}, |\phi_{\beta}\rangle\notin\mathscr{H}_\mathrm{one}$ where $|\phi_{\alpha}\rangle, |\phi_{\beta}\rangle$ are degenerate in terms of $\hat{H}_0$ with $U_{\uparrow\downarrow}=0$, but not coupled through second-order perturbation.
\begin{figure}[htb]
	\includegraphics[width =\linewidth]{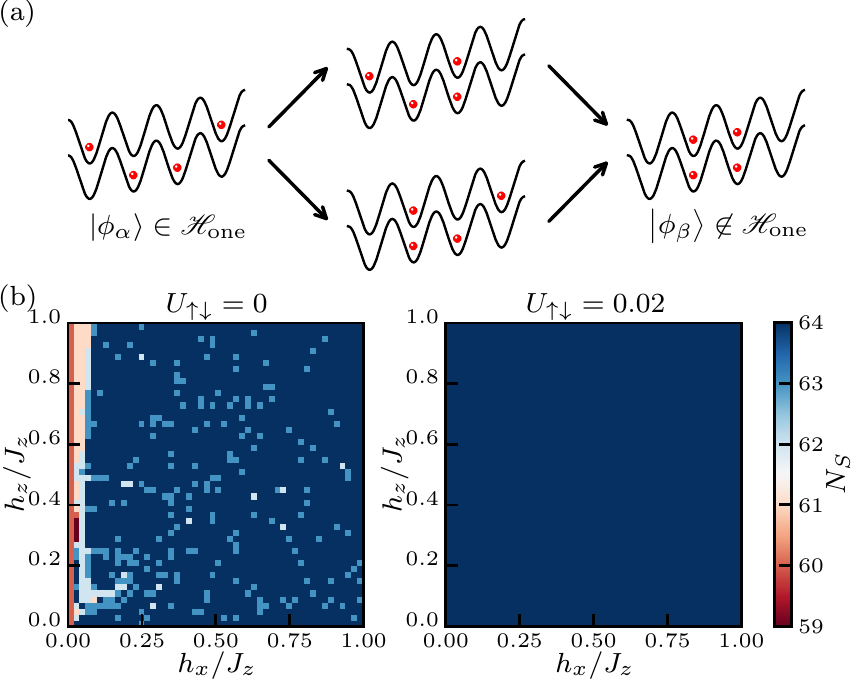}
	\caption{(a) The paired second-order resonant couplings at $U_{\uparrow\downarrow}=0$ in a four-cell lattice.  The superexchange interactions for the upper and the lower tunneling processes are $\frac{t_x^2}{\varDelta}|\phi_{\beta}\rangle\langle\phi_{\alpha}|$ and $-\frac{t_x^2}{\varDelta}|\phi_{\beta}\rangle\langle\phi_{\alpha}|$ respectively, so the total coupling between $|\phi_{\alpha}\rangle$ and $|\phi_{\beta}\rangle$ is zero.  (b) The number of selected eigenstates $N_S$ at $U_{\uparrow\downarrow}=0$ and $U_{\uparrow\downarrow}=0.02$. It can be seen that the computing error at $U_{\uparrow\downarrow}=0$ can be removed by setting a tiny nonzero $U_{\uparrow\downarrow}$. }
	%Meanwhile, it can be verified that $U_{\uparrow\downarrow}=0$ is also not a third order or higher order resonant point.
	\label{fig:figS3}
\end{figure}

\section{Parameter setting of the dynamical evolving process}
We design a process for the double-chain Bose-Hubbard model to simulate the dynamical quantum phase transition of the effective transverse Ising model from a Mott insulator which could be easily obtained in cold atom experiments. In the beginning, we assume the initial Mott insulator is prepared in thermal equilibrium in a flat lattice at a parameter $\varDelta=0, U=1, t_x=0, t_{in}=0,\delta_z=-10$. At this time, all atoms are staying on the upper chain when $k_BT\ll |\delta_z|$. The initial density operator $\hat{\rho}(0)=\operatorname{tr}(e^{-\hat{H}(0)/k_BT}/Z)$ is set by the initial temperaure derived from a designated total entropy. 

\begin{figure}[htb]
	\includegraphics[width =\linewidth]{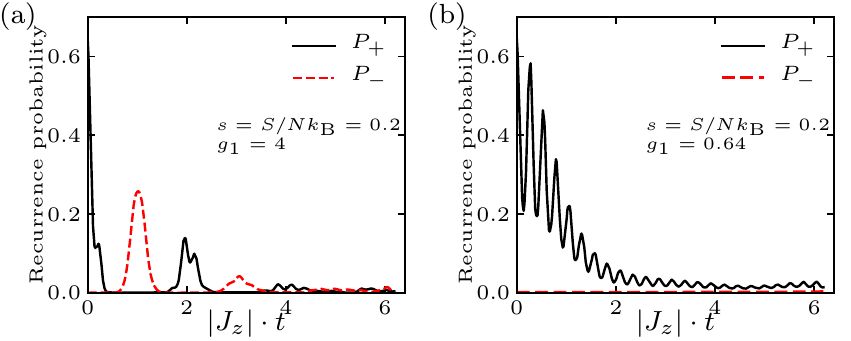}
	\caption{(a) The recurrence probability $P_+(t)$ and $P_-(t)$ after quenching from $g_0=0$ to $g_1=4$ where $g=2|h_x/J_z|$. The entropy of the initial Mott insulator is $s=S/Nk_B=0.2$. The parameters before quenching are $U=1, \varDelta=0.7, t_x=0.04, \delta_z=0, t_{in}=|J_z|=0.025$. It can be seen that there is a periodic transfer between $P_+(t)$ and $P_-(t)$ when quenching across the quantum phase transition point $g_1>g_c=1$. Here $g_c$ is the quantum phase transition point of the 1D transverse Ising model. (b)  The recurrence probability $P_+(t)$ and $P_-(t)$ of quenching from $g_0=0$ to $g_1=0.64$. There is no periodic transfer between $P_+(t)$ and $P_-(t)$ when $g_1<g_c$.}
	\label{fig:figS4}
\end{figure}

Then $\varDelta$ is increased to $\varDelta=0.7$ and $\delta_z$ is lowered to $\delta_z=0$. Next, $t_x$ is increased to $t_x=0.04$ rapidly followed by quenching of $t_{in}$. For simplicity of calculation, we assume varrying of $\varDelta, \delta_z$, $t_x$ and quenching of $t_{in}$ are all so quick that $\hat{\rho}(t_i)\approx\hat{\rho}(0)$ where $t_i$ is the time point of quenching. Basically it is enough that the varying time is much shorter than the tunneling time $\hbar/t_x$ and $\hbar/t_{in}$, keeping $\hat{\rho}_{one}(t_i)\propto|+\rangle\langle+|$. After quenching, the Hamiltonian does not change any more and controls the evolution of $\hat{\rho}(t)$ by $i\frac{\partial\hat{\rho}}{\partial t}=[\hat{H},\hat{\rho}]$. Then $P_{+}(t)=\langle+|\hat{\rho}(t)|+\rangle$ and $P_{-}(t)=\langle-|\hat{\rho}(t)|-\rangle$ can be derived, as shown in Fig. S4. It can be seen that there will be a periodic transfer between $P_+(t)$ and $P_-(t)$ when quenching across the quantum phase transition point $g_1>g_c$. Otherwise there is no periodic transfer between $P_+(t)$ and $P_-(t)$. In above deduction, we have assumed there is no obvious heating and various forms of noise within the time period of evolving process.

\end{document}